# A Practical Proposal for State Estimation at Balanced, Radial Distribution Systems

Panayiotis (Panos) Moutis
DEPsys SA
Route du Verney 20B,
1070 Puidoux, Switzerland
+41 21 546 23 00
panayiotis.moutis@depsys.ch

Omid Alizadeh-Mousavi
DEPsys SA
Route du Verney 20B,
1070 Puidoux, Switzerland
+41 21 546 23 00
omid.mousavi@depsys.ch

*Abstract*—The ever-increasing deployment of distributed resources and the opportunities offered to loads for more active roles has changed the previously unidirectional and relatively straight-forward operating profile of distribution systems (DS). DS will be required to be monitored closely for robustness and sufficient power quality. State estimation of transmission systems has consistently served as a monitoring tool, which drives system-wide control actions and, thus, ensures the operational integrity of the electric grid. An update to the classic state estimation for the case of DS is offered in this work, based on a power flow formulation for radial networks that does not require measurements or estimate of the voltage angles.

*Keywords—distribution systems, distributed generation, PMU, state estimation.*

## I. Introduction

Market incentives have been offered to promote the wide deployment of renewable energy sources [1]. Various types and sizes of Distributed Generation (DG) units [2] have been installed mostly at the Distribution System (DS) levels thanks to various subsidizing mechanisms [3]. The roll-back of the latter [4] will push DG into energy and ancillary services markets. Also, Storage Systems (either complementary to DG or as stand-alone projects) and loads enabled through demand response programs [5] already gain revenues in the said markets [6]. These points describe a landscape of a multitude of actors with diverse characteristics, different interests, which operate at a part of the grid that is barely monitored (if not at all), although it serves almost all customers in this business (in the traditional sense), the loads. That said, DS are exposed to many power quality and reliability issues [7], with no effective oversight that can respond to events, improve customer service and experience, or prevent efficiently any of these issues.

Distribution System Operators (DSO) could control and drive all the aforementioned actors in certain behaviors that will optimize DS operation [8], but the stochastic nature of the latter (especially renewables [9]), renders such an approach very challenging. Close and advanced control of DS seems to be inevitable and recent DSO projects prove that. Vast numbers of typical load energy meters are replaced with smart ones and DS transformer operation is monitored online [10]. Nevertheless, these applications do not focus on the system status, hence, fail to capture failures, errors, system disconnects, and power quality matters, let alone prevent or limit them. To answer this challenge R&D on State Estimation (SE) at DS has been gaining momentum with studies, publications and research initiatives [10-17]. SE is the algorithmic and hardware framework, which, based on the network topology of a power system (admittance matrix of the lines, transformers, etc), processes online measurements from various points in that system to calculate an as close to real-time as possible power flow solution/representation [18].

Despites the numerous works on SE at DS, only a recent few have taken a more applied approach [19, 20]. These works gather valuable results and propose a hardware design for a micro Phase Measurement Unit (uPMU). The uPMU design and the overlying SE describe a framework which is cost intensive both in capital and operating expenses.

A novel approach for SE at DS-level is here proposed. The proposal considers the monitoring requirements and operation focus of DSO. In this sense, the traditional SE is modified to account for the proper measurements and a state that best serves DS. In Section II, by referring to practices of SE at the transmission systems, the preferred characteristics of measurement devices and the SE methodology are suggested. Section III details the mathematical framework emerging from the above analysis. Tests on the proposed SE at DS are presented and commented in Section IV, while Section V describes the future of this research project.

## II. Reconsidering Traditional State Estimation Principles for the Case of Distribution Systems

### A. Measurement Requirements and Accuracy

Recently and widely, PMUs have been preferred (over SCADA) as the measuring devices on which transmission-level SE is based [21]. Their accuracy, refresh rates, and siting across monitored power systems have been thoroughly studied and standardized [16,22-26]. However, some of their features relevant to their deployment at transmission systems cannot be justified as requirements for SE at DS-level.

An error metric standardized by [21], in order to ensure efficient measurement performance by a PMU, is the Total Vector Error (TVE) of the measured signal. TVE is required to be limited to ±1% of the measured signal compared to the actual one. TVE is, thus, accounting for the accumulated

This project has received funding from the European Union's Horizon 2020 research and innovation programme under the Marie Sklodowska-Curie grant agreement No 797451.
**Preprint under Green Open Access policy**

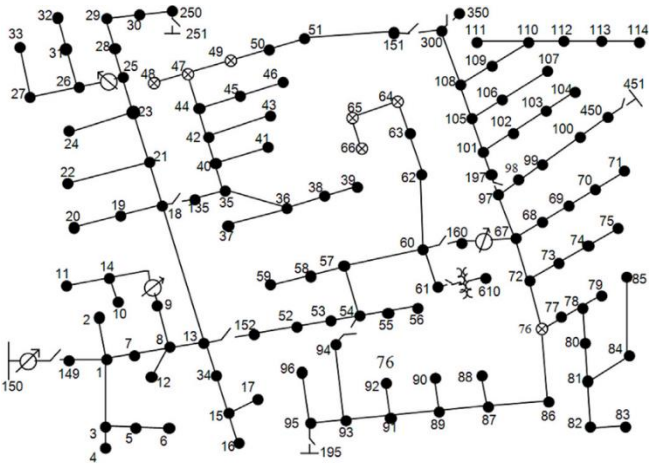

Figure 1. One-line diagram of the IEEE 123-node test distribution feeder.

TABLE I. AVERAGE ERROR OF VOLTAGE MAGNITUDE STATE ESTIMATION WHEN PREFERRING NODAL (PSEUDO-)MEASUREMENTS

| $e_i$ \ $e_v$ | 0.1% | 0.3% | 0.6% | 1% |
|---|---|---|---|---|
| 0.1% | 0.042% | 0.126% | 0.253% | 0.417% |
| 0.3% | 0.042% | 0.127% | 0.253% | 0.421% |
| 0.6% | 0.042% | 0.126% | 0.252% | 0.425% |
| 1% | 0.042% | 0.127% | 0.255% | 0.420% |

TABLE II. AVERAGE ERROR OF POWER FLOWS STATE ESTIMATION WHEN PREFERRING NODAL (PSEUDO-)MEASUREMENTS

| $e_i$ \ $e_v$ | 0.1% | 0.3% | 0.6% | 1% |
|---|---|---|---|---|
| 0.1% | 0.162% | 0.266% | 0.414% | 0.717% |
| 0.3% | 0.248% | 0.314% | 0.508% | 0.710% |
| 0.6% | 0.392% | 0.446% | 0.658% | 0.803% |
| 1% | 0.596% | 0.646% | 0.740% | 0.891% |

errors of phase angle, frequency and magnitude measurement of the signal. For the case of DS, the TVE over voltage could be required to be measured with an even smaller error bound, because relevant power quality standards allow for a ±10% voltage deviation (or less) around nominal value [27, 28]. Assuming frequency and phase angle are measured perfectly, the required TVE interval represents 1/10$^{th}$ of the power quality interval and can have ambiguous effects to the voltage regulation employed by DSO or from transformers equipped with on-load tap changers [29].

Further to the above, for DS SE, TVE should be replaced by a signal magnitude estimate error, since accounting for signal frequency or phase angle is not critical to DSO. As for frequency, DSO have no obligation in responding to events relevant to frequency deviations. Although recent standards describe more active roles for DG units in these service [30] and previous publications have established such capabilities [9,31], no market frameworks or grid codes describe a setting in which DG units can be actively involved in primary frequency control [32], hence, their behaviors are not expected to be supervised by DSO.

As for signal phase angles, measuring them at the DS-level might be obsolete. The vast majority of DS are operated radially [33], hence the voltage magnitude (compared between/among buses) is an adequate measure of whether power is *persistently* injected or absorbed at any given point or part of a DS. On the other hand, the uPMUs developed specifically for SE in DS with particularly high angle measurement precision (due to the small differences between buses), come at very high capital costs [19].

The above analysis allows for a SE framework that can omit phase angle measurements and estimation, while it requires a closer look to voltage magnitude measurement accuracy and its effect on the SE results.

*B. Measurement and Reporting Rates as IT Concerns*

Traditionally SE data are gathered and organized online and at high refresh rates [34], because phenomena of dynamic and/or transient nature need to be monitored and handled before they cause instability or market irregularities. At the DS-level though, most phenomena are not dynamic. Voltage deviations from nominal values, disconnects of part(s) of a DS, and activation of DSO equipment affect DS at the time scales of minutes and are handled in similar time frames [27].

To this end, high estimate refresh rates are an excessive requirement that can be moderated. It is interesting to note that a recent R&D effort in SE for DS has highlighted the high telecommunications costs incurred by daily loads of 0.5 GB of transmitted information over cellular networks for realizing an online SE approach [19].

As of these remarks a SE framework for DS could be assessed at the time frames of minutes and by standards of averaged values as those of power quality [27, 28].

III. PROPOSED STATE ESTIMATION FOR DISTRIBUTION SYSTEMS

State estimation is an optimization problem described as:

$$\text{minimize} \quad |z - h(x)|$$

Where, $z$ is the measurement vector and $h(x)$ the measurement function of the system state x. As of the analysis in the previous Section and based on the power flow equations also used in [35], functions $h(x)$ will be as:

$v_{x,i}^2 = v_{b,i}^2$ (1)

$v_i^2 = v_j^2 - 2 \cdot p_{ij} \cdot r_{ij} - 2 \cdot q_{ij} \cdot x_{ij}$ (2)

$p_{b,i} = \sum p_{x,ij}$ (3a)

$q_{b,i} = \sum q_{x,ij}$ (3b)

$p_{b,ij} = p_{x,ij}$ (4a)

$q_{b,ij} = q_{x,ij}$ (4b)

$p_{b,ij} = -p_{x,ji}$ (5a)

$q_{b,ij} = -q_{x,ji}$ (5b)

Where subscripts $b$ and $x$ denote measurement and state of the variable, respectively. (1) & (4) describe the estimated equality between measurement and state for the voltage magnitude and the power flow, respectively. (2) describes the voltage drop along line $ij$, where any one of the flows or voltages can be replaced by a measurement and the rest will be state variables. (3) is the power balance of flows in/out of a node. (5) is the expected "equality" (reminding that the formulation corresponds to a linear approximation) of the two opposite direction flows along line $ij$. As it may be seen, the voltage angles are not considered in this power flow set-up. According to the optimal power flow study conducted in [35], this set-up boasts low errors in the calculation of voltage magnitudes, which is expected to reflect in the here proposed state estimation formulation.

TABLE III. AVERAGE MAXIMUM ERROR OF VOLTAGE MAGNITUDE STATE ESTIMATION WHEN PREFERRING NODAL (PSEUDO-)MEASUREMENTS

| $e_i$ \ $e_v$ | 0.1% | 0.3% | 0.6% | 1% |
|---|---|---|---|---|
| 0.1% | 0.134% | 0.396% | 0.793% | 1.313% |
| 0.3% | 0.132% | 0.400% | 0.797% | 1.325% |
| 0.6% | 0.133% | 0.401% | 0.793% | 1.333% |
| 1% | 0.132% | 0.401% | 0.805% | 1.318% |

TABLE IV. AVERAGE MAXIMUM ERROR OF POWER FLOWS STATE ESTIMATION WHEN PREFERRING NODAL (PSEUDO-)MEASUREMENTS

| $e_i$ \ $e_v$ | 0.1% | 0.3% | 0.6% | 1% |
|---|---|---|---|---|
| 0.1% | 6.957% | 8.084% | 7.731% | 15.861% |
| 0.3% | 7.667% | 8.364% | 12.878% | 13.549% |
| 0.6% | 8.300% | 9.587% | 17.001% | 17.094% |
| 1% | 9.943% | 11.815% | 14.199% | 15.340% |

TABLE V. AVERAGE ERROR OF VOLTAGE MAGNITUDE STATE ESTIMATION WHEN PREFERRING EDGE (PSEUDO-)MEASUREMENTS

| $e_i$ \ $e_v$ | 0.1% | 0.3% | 0.6% | 1% |
|---|---|---|---|---|
| 0.1% | 0.053% | 0.158% | 0.319% | 0.524% |
| 0.3% | 0.053% | 0.161% | 0.317% | 0.532% |
| 0.6% | 0.053% | 0.157% | 0.318% | 0.530% |
| 1% | 0.053% | 0.159% | 0.315% | 0.530% |

TABLE VI. AVERAGE ERROR OF POWER FLOWS STATE ESTIMATION WHEN PREFERRING EDGE (PSEUDO-)MEASUREMENTS

| $e_i$ \ $e_v$ | 0.1% | 0.3% | 0.6% | 1% |
|---|---|---|---|---|
| 0.1% | 0.154% | 0.241% | 0.400% | 0.626% |
| 0.3% | 0.237% | 0.295% | 0.437% | 0.652% |
| 0.6% | 0.397% | 0.432% | 0.536% | 0.716% |
| 1% | 0.618% | 0.640% | 0.715% | 0.865% |

TABLE VII. AVERAGE MAXIMUM ERROR OF VOLTAGE MAGNITUDE STATE ESTIMATION WHEN PREFERRING EDGE (PSEUDO-)MEASUREMENTS

| $e_i$ \ $e_v$ | 0.1% | 0.3% | 0.6% | 1% |
|---|---|---|---|---|
| 0.1% | 0.143% | 0.431% | 0.864% | 1.401% |
| 0.3% | 0.142% | 0.434% | 0.850% | 1.445% |
| 0.6% | 0.143% | 0.427% | 0.861% | 1.432% |
| 1% | 0.143% | 0.433% | 0.852% | 1.435% |

TABLE VIII. AVERAGE MAXIMUM ERROR OF POWER FLOWS STATE ESTIMATION WHEN PREFERRING NODAL (PSEUDO-)MEASUREMENTS

| $e_i$ \ $e_v$ | 0.1% | 0.3% | 0.6% | 1% |
|---|---|---|---|---|
| 0.1% | 3.711% | 4.448% | 6.248% | 9.433% |
| 0.3% | 4.384% | 5.117% | 6.946% | 9.806% |
| 0.6% | 6.410% | 6.842% | 8.142% | 10.748% |
| 1% | 9.410% | 9.567% | 10.728% | 12.959% |

## IV. PROPOSED STATE ESTIMATION PERFORMANCE ASSESSMENT

### A. Testing set-up

The IEEE-123 test system (Fig. 1) is used to assess the performance of the proposed SE at DS. Besides its typical characteristics, 49 DG units are assumed to have been installed as described in the Appendix. The base power set was 5MVA. By varying the cost of the DG units various possible dispatches are generated, hence various voltage profiles and power flows will occur in the DS. As of the analysis in Section II the next parameters define the testing scenarios:

i. Voltage magnitude measurement error $e_v$ of 0.1%, 0.3%, 0.6%, 1% compared to the actual signal,

ii. Current magnitude measurement error $e_i$ of 0.1%, 0.3%, 0.6%, 1% compared to the actual signal,

iii. Measurements of line flows may be preferred over measurements of bus voltage and injected/absorbed active and reactive powers and vice versa.

Regarding parameter (iii), it is assumed that for every different dispatching of the DG units, there exists a different (randomized) set of measurements and/or pseudo-measurements. When referring to *nodal measurements* in this study, measurement of active and reactive injected/absorbed powers and voltage magnitudes for a node/bus are included. When referring to *edge measurements* in this study, measurement of active and reactive power flows flowing from (positive) or to (negative) one node are considered.

The measurements are such that the system is observable. The reason for considering this parameter is that different sets of pseudo-measurements might be used to complement the actual measurements, thus offering an indication of the effect of nodal over edge (pseudo-)measurements. The number of estimated bus voltages and line flows are logged for every state estimation test. It is noted that there are 61 load/generation buses and 60 lines in this test-system.

As of all possible combinations among the above three parameters, 32 scenarios are determined. For each scenario 1500 random dispatches were tested. From each dispatch the average and maximum errors of the estimate of bus voltages and line power flows compared to their actual values as from the exact power flow solution are recorded. The *average estimate error* over one execution of the estimate method for a tested dispatch and as of the definitions in (1)-(5) is as:

$$\widetilde{\Delta x} = \frac{|x - m|}{\widetilde{m}}$$

Where $x$ and $m$ are considered as stacked vectors of the state variables estimated and their actual values, respectively. The average of those averages for every set of tests run for each scenario as of the combinations in IV.A.(i-iii) are presented in Tables I, II, V, VI. The *maximum estimate error* over one execution of the estimate method for a tested dispatch and as of the definitions in (1)-(5) is as:

$$\max\left(\frac{|x - m|}{\widetilde{m}}\right)$$

The average of those maxima for every set of tests run for each scenario as of the combinations in IV.A.(i-iii) are presented in Tables III, IV, VII, VIII. The average of the maximum errors is an indication of whether there are persistent high errors for particular types (subsets) of test cases.

### B. Results and discussion

For the case of nodal (pseudo-)measurements preferred over edge ones, the average estimate error for voltage magnitudes and power flows are given in Tables I and II for the various measurements errors. The average of maximum estimate error from each test are given in Tables III and IV. The same statistics for the case of edge (pseudo-)measurements preferred over nodal ones are summarized in Tables V-VIII.

As from the results the following remarks can be made:

i. The current measurement error does not affect almost at all the estimate error of the voltage magnitudes. This can be explained by the fact that the underlying formulation performs very positively in the approximation of voltage magnitudes [35].
ii. The measurement errors of voltage and current have an equal effect on the estimate error of the power flows, which is expected as of definition of active/reactive power and the subsequent error propagation [36].
iii. The average of the maximum errors of the voltage magnitude estimation is well within the interval of power quality limits [27, 28]. The reasoning in (i) may be repeated here.
iv. The maximum errors of power flow estimation are lower for the case that edge (pseudo-)measurements are preferred over nodal measurements. All other statistics remain unaffected by this preference.

Especially for the case of power flow estimation errors, further analysis showed that extreme errors that would affect the corresponding averages were limited to one or two flows (active or reactive) out of the 240 (120 pairs of active and reactive), while the ones in the opposite direction would be properly estimated. It is easy to mitigate such outliers by a post-processing filtering.

As for the required measurements and pseudo-measurements, in order to make the system observable, it is very interesting to note that if nodal measurements are preferred over edge ones, then 60% of nodes and 80% of the flows need to serve as input, while for the inverse case the required percentages are 30% of nodes and 90% of flows. In other words, by a small increase in the (pseudo-)measurements of flows, there is a significant improvement in the output of voltage magnitude estimation.

## V. CONCLUSIONS

In this work the typical SE was rethought based on valid concerns arising with the monitoring of DS. Small voltage phase angles and narrow power quality intervals for voltage magnitudes imply that the SE framework needs to be refocused accordingly. To this end, a SE formulation that corresponds to these concerns is proposed and assessed. In principle, the methodology shows to have positive results in addressing the matters at hand; the voltage magnitude estimates are of high accuracy and by properly designing the measurement infrastructure the DS monitoring can be adequately wide. Some high errors in specific power flow estimates can be filtered beyond the SE framework itself, but some additional studies could offer valuable insights.

## APPENDIX

The voltage regulators, transformers and switches are omitted from the IEEE 123-node test distribution feeder. The DG units added are given in Table IX.

TABLE IX
DG UNITS ASSUMED CONNECTED TO BUSES OF
THE IEEE 123-NODE TEST DISTRIBUTION FEEDER FOR TESTING

| Bus No. as of IEEE-123 | $P_n$ (kW) | Bus No. as of IEEE-123 | $P_n$ (kW) |
|---|---|---|---|
| 149 | (slack bus) | 64 | 10 |
| 1 | 40 | 65 | 50 |
| 13 | 10 | 66 | 40 |
| 23 | 40 | 67 | 40 |
| 25 | 20 | 72 | 10 |
| 28 | 10 | 76 | 10 |
| 29 | 10 | 78 | 40 |
| 30 | 10 | 79 | 30 |
| 35 | 20 | 80 | 30 |
| 40 | 10 | 86 | 50 |
| 42 | 40 | 87 | 20 |
| 44 | 10 | 89 | 10 |
| 47 | 30 | 91 | 30 |
| 48 | 20 | 93 | 40 |
| 50 | 10 | 95 | 20 |
| 51 | 50 | 97 | 10 |
| 53 | 20 | 98 | 30 |
| 54 | 30 | 99 | 20 |
| 55 | 30 | 100 | 40 |
| 56 | 10 | 105 | 30 |
| 57 | 20 | 108 | 40 |
| 60 | 20 | 151 | 30 |
| 61 | 20 | 250 | 10 |
| 62 | 20 | 300 | 10 |
| 63 | 30 | 450 | 10 |